\documentclass[aps,reprint,twocolumn,superscriptaddress]{revtex4-1}
\usepackage[english]{babel}
\usepackage{subfigure}
\usepackage{amsmath, amssymb,graphicx,color,bm,epstopdf}
\usepackage[dvipsnames]{xcolor}
\usepackage{bbold}
\usepackage{braket}
\usepackage{comment}
\usepackage{natbib}
\usepackage{slashed}
\usepackage{enumitem} 

\newcommand{\be}{\begin{equation}}
\newcommand{\e}{\end{equation}}
\newcommand{\beml}{\begin{subequations}}
\newcommand{\eml}{\end{subequations}}
\newcommand{\beq}{\begin{eqnarray}}
\newcommand{\eq}{\end{eqnarray}}
\newcommand{\ba}{\begin{array}}
\newcommand{\ea}{\end{array}}
\newcommand{\bpm}{\begin{pmatrix}}
\newcommand{\epm}{\end{pmatrix}}
\newcommand{\bc}{\begin{cases}}
\newcommand{\ec}{\end{cases}}

\usepackage{placeins}
\usepackage{float}
\usepackage{upgreek}

\definecolor{amendments}{rgb}{0.0, 0.0, 0.7}

\usepackage[utf8]{inputenc}
\begin{document}
\title{Scattering of a twisted electron wavepacket by a finite laser pulse}
\author{I.~A. Aleksandrov}
\affiliation{Department of Physics, St. Petersburg State University, 7/9 Universitetskaya Naberezhnaya, Saint Petersburg 199034, Russia}
\affiliation{Ioffe Institute, Politekhnicheskaya street 26, Saint Petersburg 194021, Russia}
\author{D.~A. Tumakov}
\affiliation{Department of Physics, St. Petersburg State University, 7/9 Universitetskaya Naberezhnaya, Saint Petersburg 199034, Russia}
\affiliation{Ioffe Institute, Politekhnicheskaya street 26, Saint Petersburg 194021, Russia}
\author{A. Kudlis}
\affiliation{ITMO University, Kronverkskiy prospekt 49, Saint Petersburg 197101, Russia}
\author{V.~A. Zaytsev}
\affiliation{Department of Physics, St. Petersburg State University, 7/9 Universitetskaya Naberezhnaya, Saint Petersburg 199034, Russia}
\author{N.~N. Rosanov}
\affiliation{Ioffe Institute, Politekhnicheskaya street 26, Saint Petersburg 194021, Russia}

\begin{abstract}
The behavior of a twisted electron colliding with a linearly polarized laser pulse is investigated within relativistic quantum mechanics. In order to better fit the real experimental conditions, we introduce a Gaussian spatial profile for the initial electron state as well as an envelope function for the laser pulse, so the both interacting objects have a finite size along the laser propagation direction. For this setup we analyze the dynamics of various observable quantities regarding the electron state: the probability density, angular momentum, and mean values of the spatial coordinates. It is shown that the motion of a twisted wavepacket can be accurately described by averaging over classical trajectories with various directions of the transverse momentum component. On the other hand, full quantum simulations demonstrate that the ring structure of the wavepacket in the transverse plane can be significantly distorted leading to large uncertainties in the total angular momentum of the electron. This effect remains after the interaction once the laser pulse has a nonzero electric-field area.

\end{abstract}

\maketitle

\section{Introduction}\label{sec:intro}

The both massive and massless particles which are capable of carrying nonzero orbital angular momentum (OAM) are of a great interest for researchers from different fields of study. The first seminal description of the concept of such \textit{twisted} particles was introduced by Allen~\textit{et al.} in the early 90s in Ref.~\cite{AllenPRA1992}, where it was discussed in the context of photons. 
The wave front of the corresponding beams has a helical spatial structure allowing the beams to carry a nonzero OAM projection onto the propagation direction. Soon afterwards, the fact that twisted photons possess discreet OAM was confirmed experimentally in Ref.~\cite{BeijersbergenOptComm1993}, and two years later He \textit{et al.} managed to transfer OAM from light to matter~\cite{PhysRevLett.75.826}. In fact, the opportunities related to the information capacity and other specificities of these \textit{optical vortices} -- twisted photons -- have encouraged numerous studies concerning a huge variety of their applications~\cite{torres2011twisted} (see also reviews~\cite{ALLEN1999291,Soskin2001219,Padgett2004,FrankeArnoldLasPhot2008,Knyazev_Serbo,allen2016}).

Despite the fact that the development of the field began with the investigation of photons, it was clear from the very beginning that massive twisted particles could also be produced. The fundamental possibility that not only photons can carry OAM but also, for example, electrons was indicated in a number of studies (see, e.g., Refs.~\cite{OuyangAnnPhys1999,BagrovAnnPhys2005,LearyNJP2008,BliokhPRL2011}). For practical applications, an interesting feature of twisted electrons is that, unlike photons, they carry a magnetic moment proportional to OAM which can experimentally reach hundreds of $\hbar$~\cite{Mafakheri_2017}. It allows electron vortices to effectively interact with external magnetic fields~\cite{BliokhPRL2007,VerbeeckNat2010,BliokhPRL2011,PhysRevLett.108.044801,PhysRevLett.108.074802,PhysRevA.86.012701,PhysRevX.2.041011}, which makes them a versatile tool for investigating magnetic properties of different materials~\cite{PhysRevLett.111.105504, Beche_NP10_26_2014, Schattschneider_U136_81_2014, PhysRevLett.116.127203} as well as for detecting such subtle magnetic effects as, e.g., polarization radiation~\cite{PhysRevLett.110.264801, PhysRevA.88.043840}.
The method proposed by Bliokh~\textit{et al.} in Ref.~\cite{BliokhPRL2007} for producing such electrons was successfully employed by several groups of experimentalists who have played a pioneering role in this field of study~\cite{UchidaNat2010,VerbeeckNat2010,McMorranSc2011}~(see also recent papers~\cite{10.1093/jmicro/dfs036,PhysRevLett.109.084801,PhysRevLett.111.064801,Ryu_2014}). 
The energy scale in these experiments amounted to $\sim 300$~keV while the orbital quantum number reached one hundred. 
This experimental breakthrough has motivated theorists to deeper analyze the properties of twisted electrons and also to propose new setups for their generation and investigation~\cite{SCHATTSCHNEIDER20111461,PhysRevLett.108.044801,PhysRevLett.108.074802,PhysRevA.86.012701,doi:10.1063/1.4863564,PhysRevLett.114.034801,serbo_pra_2015}. 
The positive outcome opened broad prospects not only for researchers in nuclear and high-energy physics, where this feature may serve as an efficient tool for a better understanding of processes occurring in collision and scattering experiments~\cite{IvanovPRD2011,IvanovPRA2011,PhysRevA.85.033813,Ivanov_2020}, but also for investigations dealing with magnetic and biological materials~\cite{VerbeeckAPL2011,SchattschneiderPRB2012,GuzzinatiPRL2013}. A detailed description of theoretical and experimental studies that have emerged over the last decade was presented by Bliokh \textit{et al.} in a comprehensive review~\cite{BLIOKH20171}. One should note here that the relativistic and nonrelativistic descriptions lead to substantially different results for observable quantities. In particular, due to the nontrivial coupling of the electron spin and its angular momentum according to the Dirac equation, the latter is not conserved in contrast to the case of the nonrelativistic treatment. The detailed analyses of Schr\"{o}dinger equation for the case of twisted electrons can be found, e.g., in Ref.~\cite{Schattschneider_2011}.

In order to describe the behavior of twisted electrons within relativistic theory, one has to construct the corresponding solutions of the Dirac equation which differ from the usual plane-wave states. For instance, the solution in the form of the so-called Bessel beams can be considered~\cite{BliokhPRL2011}. However, one should also find out whether such solutions properly mimic the real experimental setups. For example, the Bessel beams themselves are not localized in space.
To eliminate this shortcoming, one can introduce wavepackets~(WP) or consider other solutions of the Dirac equation, e.g., the so-called exponential WP proposed in Refs.~\cite{Bialynicki_2017,Bialynicki_2019}. Alternatively, relativistic electron vortices can be described by using the Foldy-Wouthuysen transformation of the Dirac equation followed by a series of quite natural approximations~\cite{Barnett_2017}. It allows one to derive the equations that can be solved much easier in contrast to the Dirac one. However, the legitimacy of the Foldy-Wouthuysen representation in this context is not that clear. In particular, the determination of the correct operators for physical observables and the possibility of a probabilistic interpretation of the wavefunction have given rise to heated debates~\cite{Silenko_2018,Bialynicki_2019_vs_belorus,Silenko_2019}. Recently, the number of approaches has been published which offer alternative methods for obtaining exact solutions of the Dirac equation on the basis of Bateman-Hillion ansatz~\cite{Ducharme_2021} and by means of newly developed relativistic dynamic inversion approach~\cite{Campos_2021}.

The list of theoretical studies of physical processes involving twisted electrons is very extensive (see, e.g., Refs.~\cite{Seipt_2014,Bandyopadhyay_2015,Matula_2014,KarlovetsPRA2017,Zaytsev_2017,Kosheleva_2018,Groshev_2020,KarlovetsPRA2012,HayrapetyanPRL2014}). Let us note that the formalism describing the phenomena associated with the scattering processes of electrons in matter is relatively well developed; the recent studies have covered a number of important issues playing a vital role for the completeness of the theoretical basis~\cite{Matula_2014,KarlovetsPRA2017,Zaytsev_2017,Kosheleva_2018,Groshev_2020}. This contrasts, however, with the study of the behavior of twisted electrons colliding with a laser pulse. Although the problem and related issues have already been considered in a number of works~\cite{KarlovetsPRA2012,HayrapetyanPRL2014}, we aim to obtain more accurate results by taking into account the finite size of both the electron WP and the laser pulse. For this purpose, within relativistic quantum mechanics, we follow the theoretical approach employed in Ref.~\cite{Aleksandrov_2020} for a plane-wave electron, where the spatial envelope of the laser pulse along its propagation direction was introduced and the initial electron WP had a Gaussian profile. Following this scheme, one explores a more realistic setup from the experimental viewpoint.

Our computations are based on the following procedure. First, a localized electron WP is constructed from the Bessel beams while the laser pulse is assumed to be a linearly polarized plane wave. The WP is then expanded in terms of the Volkov solutions~\cite{Wolkow1935}. Having obtained the corresponding expansion coefficients, one can construct the solution of the Dirac equation at arbitrary time. We analyze various observable quantities such as the probability density, angular momentum and its dispersion, and mean values of the coordinates. These quantities are computed for various parameters of the external laser field and the initial electron state.

The paper is organized as follows. In Sec.~\ref{sec:scheme} we describe the setup under consideration. In Sec.~\ref{sec:plane_wave} we briefly outline the approach based on the expansion of the wavefunction in terms of the Volkov solutions in the case of a plane-wave electron state. Section~\ref{sec:twisted} is devoted to a detailed description of the analogous method for treating twisted electrons. We discuss our numerical results in Sec.~\ref{sec:results}.

Throughout the paper, we use atomic units $\hbar = m_e = -e = 1$.

\section{Setup. Electron wavepacket colliding with a laser pulse}\label{sec:scheme}
\begin{figure}[b]
\center{\includegraphics[width=0.98\linewidth]{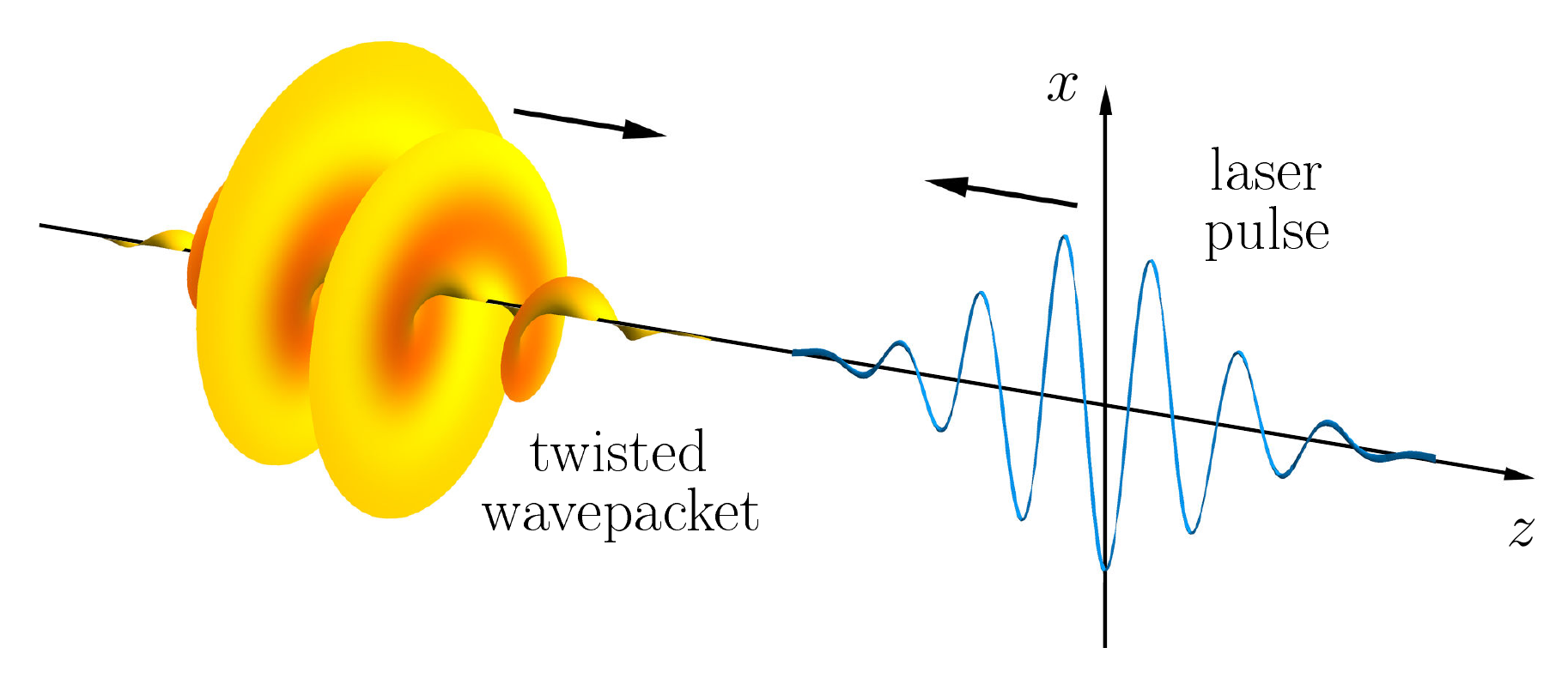}}
\caption{Initial state of the counterpropagating twisted wavepacket (left) and laser pulse (right). The momentum projection of the former is $p_z \equiv p_\parallel$. The laser field is polarized along the $x$ direction according to Eq.~(\ref{eq:field_Ex}) and has amplitude $E_*$. The center of the initial WP corresponds to $z=0$ (the $x$ axis in the figure does not cross the origin).}
\label{fig:scheme}
\end{figure}

We consider a twisted electron wavepacket propagating along the $z$ axis and a linearly polarized plane-wave laser pulse traveling along the opposite direction (see Fig.~\ref{fig:scheme}). 
The latter is described by the following expression for the electric component:
\begin{eqnarray}
&&E_x (t, z) = \mathcal{E}(ct + z), \label{eq:field_Ex} \\
&&\mathcal{E}(\xi) = E_* F(\omega \xi / c) \sin (\omega \xi /c  + \phi).
\label{eq:field}
\end{eqnarray}
Here $E_*$ and $\omega$ are the field amplitude and frequency, respectively, the function $F$ is a smooth Gaussian-shape envelope, $F(\eta) = \mathrm{exp} (-\eta^2/a^2)$, where $a$ is a dimensionless parameter governing the pulse duration, and $\phi$ is a carrier-envelope phase (CEP). It is convinient to introduce the field peak intensity
\begin{equation}
I = \frac{1}{8 \pi \alpha} E_{*}^2,
\end{equation}
where $\alpha$ is the finite structure constant.
The only nonzero component of the potential $A^\mu$ is $A^1 = A_x (t, z) = \mathcal{A}(ct + z)$ with
\begin{equation}
\mathcal{A}(\xi) = - c \int \limits_{-\infty}^{\xi} \mathcal{E} (\xi') d\xi'.
\label{eq:potential_xi}
\end{equation}
We underline here that the vector potential does not necessarily vanish for $\xi \to +\infty$. We will denote this limit by $\mathcal{A}_0 = \mathcal{A}(+\infty)$. Nonzero values of $\mathcal{A}_0$ correspond to a nonzero unipolarity of the laser pulse, i.e., nonzero momentum transferred to the electron along the electric field direction:
\begin{equation}\label{eq:field_area}
\mathcal{A}_0 =  - c \int \limits_{-\infty}^{+\infty} \mathcal{E} (\xi') d\xi' = - \frac{\sqrt{\pi} c a E_*}{\omega} \, \mathrm{e}^{- a^2 / 4} \sin \phi.
\end{equation}

In what follows, we will exploit the fact that the envelope $F(\eta)$ has an almost finite support, i.e., the function $\mathcal{A} (\xi)$ almost vanishes for $\xi < -\xi_\text{max}$ and does not differ much from $\mathcal{A}_0$ for $\xi > \xi_\text{max}$. This means that at the initial time instant $t_\text{in} = -(L+ \xi_\text{max})/c$, the laser pulse is localized within the region $z \in (L,\ L + 2\xi_\text{max})$, where $L$ should be sufficiently large, so that the pulse and the electron WP do not overlap. 

The final state after the interaction is considered at $t = t_\text{out} = (\tilde{L} + \xi_\text{max})/c$, where $-\tilde{L}$ is the position of the right edge of the laser pulse. One has to make sure that $\tilde{L}$ is sufficiently large, so that the external field and the electron WP no longer overlap. 

In what follows, we will discuss how the initial state of the electron WP can be evolved in time by means of the exact (Volkov) solutions of the Dirac equation in the external plane-wave background.

\section{Plane-wave electron state}\label{sec:plane_wave}

First, we will examine a plane-wave electron state and briefly recap the approach which we previously employed in Ref.~\cite{Aleksandrov_2020}. Since the external field does not depend on $x$ and $y$, the corresponding components of the generalized momentum are conserved. With respect to the $z$ direction, we construct a WP of the following form:
\begin{equation}
\psi^\text{(0)}_{\boldsymbol{p}, s} (\boldsymbol{x}) = \frac{1}{(2 \pi)^{3/2}} \, \mathrm{e}^{i \boldsymbol{p} \boldsymbol{x}} \! \int \limits_{-\infty}^{+\infty} \! dq \, f(q) \, \mathrm{e}^{iqz} u (\boldsymbol{p} + q \boldsymbol{n}, s),
\label{eq:psi_init}
\end{equation}
where $\boldsymbol{n}$ is a unit vector along $z$ and the smearing function $f(q)$ determines the spectral content of the initial WP. It is normalized to ensure $\langle \psi^\text{(0)}_{\boldsymbol{p}, s} | \psi^\text{(0)}_{\boldsymbol{p}, s'} \rangle = \delta_{s s'}$. The initial spin state of the electron is governed by the constant bispinors $u (\boldsymbol{p}, s)$ corresponding to the positive-energy solutions of the Dirac equation ($s = \pm$). 
Together with bispinors $v (\boldsymbol{p}, s)$, they form an orthonormal and complete set. These bispinors obey
\begin{eqnarray}
\big ( c \, \boldsymbol{\alpha} \cdot \boldsymbol{p} + \beta c^2 \big ) u (\boldsymbol{p},s) &=& \varepsilon u (\boldsymbol{p},s), \label{eq:de_u}\\
\big ( c \, \boldsymbol{\alpha} \cdot \boldsymbol{p} + \beta c^2 \big ) v (\boldsymbol{p},s) &=& -\varepsilon v (\boldsymbol{p},s), \label{eq:de_v}
\end{eqnarray}
where $\alpha_i$ and $\beta$ are the standard Dirac matrices, and $\varepsilon = c \sqrt{c^2 + \boldsymbol{p}^2}$.

The initial condition is $\psi_{\boldsymbol{p}, s} (t_\text{in}, \boldsymbol{x}) = \psi^\text{(0)}_{\boldsymbol{p}, s} (\boldsymbol{x})$, and our goal is to evolve the wavefunction in time. The main idea is to expand the initial state in terms of the Volkov solutions~\cite{Wolkow1935} and then combine them at some given $t$ using the same coefficients as they do not depend on time. The Volkov states are given by
\begin{widetext}
\begin{eqnarray}
\varphi^{(\zeta)}_{\boldsymbol{p}', s'} (t, \boldsymbol{x}) &=& \frac{1}{(2 \pi)^{3/2}} \, \mathrm{e}^{i \zeta \boldsymbol{p}' \boldsymbol{x}} f^{(\zeta)}_{\boldsymbol{p}', s'} (t, z),\label{eq:volkov_1}\\
f^{(\zeta)}_{\boldsymbol{p}', s'} (t, z) &=& \mathrm{e}^{- i \zeta \varepsilon' t}  \, \mathrm{exp} \Bigg \{ i \int \limits_{-\infty}^{n \cdot x} \! d\xi \, \frac{1}{2 (n \cdot p')} \, \bigg [ \frac{2}{c} \, (p' \cdot A(\xi)) + \frac{\zeta}{c^2} \, A^2(\xi) \bigg ] \Bigg \} \notag \\
{} &\times & \bigg [ 1 - \frac{\zeta}{2c (n \cdot p')} \, (\gamma \cdot n) (\gamma \cdot A) \bigg ] w_\zeta (\boldsymbol{p}', s'),\label{eq:volkov_2}
\end{eqnarray}
where $\zeta = \pm$ denotes the sign of the energy, $w_+ (\boldsymbol{p}', s') = u (\boldsymbol{p}', s')$, and $w_- (\boldsymbol{p}', s') = v (-\boldsymbol{p}', s')$. In our case, it yields
\begin{eqnarray}
f^{(\zeta)}_{\boldsymbol{p}', s'} (t, z) &=& \mathrm{e}^{- i \zeta \varepsilon' t}  \, \mathrm{exp} \Bigg \{ -\frac{i}{\varepsilon' + c p_z'} \Bigg [ p_x' \int \limits_{-\infty}^{\xi} \! d\xi' \, \mathcal{A} (\xi') + \frac{\zeta}{2c} \int \limits_{-\infty}^{\xi} \! d\xi' \, \mathcal{A}^2 (\xi') \Bigg ] \Bigg \} \notag \\
{} &\times & \bigg [ 1 + \frac{\zeta}{2 (\varepsilon' + c p_z')} \, \mathcal{A} (\xi) (\gamma^0 + \gamma^3) \gamma^1 \bigg ] w_\zeta (\boldsymbol{p}', s'),
\label{eq:volkov_f_specified}
\end{eqnarray}
where $\xi = n\cdot x = ct + z$.
\end{widetext}

To construct the time-dependent solution of the Dirac equation, we make use of the fact that the Volkov states form an orthonormal and complete set~\cite{boca2010, dipiazza2018}, so the wavefunction can be expanded as follows:
\begin{equation}
\psi_{\boldsymbol{p}, s} (t, \boldsymbol{x}) = \sum_{\zeta} \sum_{s'} \int \! d\boldsymbol{p}' \, C^{(\zeta)}_{\boldsymbol{p}', s'} \varphi^{(\zeta)}_{\boldsymbol{p}', s'} (t, \boldsymbol{x}),
\end{equation}
where the coefficients $C^{(\zeta)}_{\boldsymbol{p}', s'}$ do not depend on time because the Dirac Hamiltonian is Hermitian. They can be evaluated at the time instant $t = t_\text{in}$ as a usual inner product:
\begin{eqnarray}
C^{(\zeta)}_{\boldsymbol{p}', s'} &=& \langle \varphi^{(\zeta)}_{\boldsymbol{p}', s'} (t_\text{in}) | \psi^\text{(0)}_{\boldsymbol{p}, s} \rangle \nonumber \\
{} &\equiv& \int \! d\boldsymbol{x} \, \big [\varphi^{(\zeta)}_{\boldsymbol{p}', s'} (t_\text{in}, \boldsymbol{x}) \big ]^\dagger \psi^\text{(0)}_{\boldsymbol{p}, s} (\boldsymbol{x}).
\label{eq:coeff_C}
\end{eqnarray}
It turns out that in the plane-wave background, these coefficients are ``diagonal'' with respect to $p_x$ and $p_y$: $C^{(\zeta)}_{\boldsymbol{p}', s'} = \delta (p_x' - \zeta p_x) \delta (p_y' - \zeta p_y) c^{(\zeta)}_{p_z', s'}$. One can explicitly verify that $-i \partial_x \varphi^{(+)}_{\boldsymbol{p}', s'} (t, \boldsymbol{x}) = p_x' \varphi^{(+)}_{\boldsymbol{p}', s'} (t, \boldsymbol{x})$, so the index $p_x'$ corresponds to the generalized momentum. Since the initial wavefunction~\eqref{eq:psi_init} corresponds to the positive-energy subspace of solutions, the coefficients $c^{(-)}_{\boldsymbol{p}', s'}$ vanish, so one can use only the positive-energy Volkov solutions ($\zeta = +$). Note that the Volkov functions possess a well-defined sign of the energy which does not depend on time (this is in accordance with the fact that a plane-wave background does not produce $e^+e^-$ pairs). We obtain
\begin{eqnarray}
c^{(+)}_{p_z', s'} &=& \int \limits_{-\infty}^{+\infty} \! \frac{dz}{2 \pi} \int \limits_{-\infty}^{+\infty} \! dq \, f(q) \, \mathrm{e}^{i (p_z - p_z')z} \mathrm{e}^{iqz} \nonumber \\
{}&\times& \big [ f^{(+)}_{\boldsymbol{p}', s'} (t_\text{in}, z) \big ]^\dagger u (\boldsymbol{p} + q \boldsymbol{n}, s),
\label{eq:coeff}
\end{eqnarray}
where $p'_x = p_x$ and $p_y' = p_y$. The wavefunction can be then evaluated at arbitrary time instant $t$ via
\begin{equation}
\psi_{\boldsymbol{p}, s} (t, \boldsymbol{x}) = \sum_{s'} \int \! dp_z' \, c^{(+)}_{p_z', s'} \varphi^{(+)}_{p_x, p_y, p_z', s'} (t, \boldsymbol{x}).
\label{eq:psi_c_t}
\end{equation}

\begin{widetext}

\section{Twisted electron}\label{sec:twisted}

In this section, we will discuss how the approach described above should alter if one replaces a plane-wave electron with a twisted one. First, we discuss the case of Bessel beams, which are spatially infinite, and then turn to the analysis of localized WPs.

\subsection{Bessel beam}\label{sec:bessel_beam}
Let us consider the so-called Bessel beam of twisted electrons traveling along the $z$ axis ($p_\parallel \equiv p_z$)~\cite{BliokhPRL2011}:
\begin{equation}\label{eq:bessel_init}
    \psi^{(\text{B})}_{l,p_{\parallel},p_{\perp},s}(\boldsymbol{x})=\frac{\sqrt{p_\perp}}{2^{3/2} \pi} \, \mathrm{e}^{ip_{\parallel}z} g_{l,p_{\parallel},p_{\perp},s}(x,y),
\end{equation}
where
\begin{equation}
    g_{l,p_{\parallel},p_{\perp},s}(x,y) \equiv g_{l,p_{\parallel},p_{\perp},s}(\rho, \varphi) =
    \sum_{k=-1}^{+1} i^k a_k \mathrm{e}^{i(l+k)\varphi} J_{l+k} (p_\perp \rho).
    \label{eq:g_AAA}
\end{equation}
Here
 \begin{equation}
    a_0 =  
     \left(\begin{tabular}{c}
     {$\sqrt{1+c^2/\varepsilon} \ \vec{w}$}\\
     \\
     {$\sqrt{1-c^2/\varepsilon}\ \sigma_z \cos \theta_0 \, \vec{w}$}  \\
     \\
     \end{tabular}\right), \qquad 
    a_{-1} =\begin{pmatrix}
           0 \\
           0 \\
           \bar{\beta} \sqrt{\Delta} \\
           0
     \end{pmatrix}, \qquad
     a_{1} =\begin{pmatrix}
           0 \\
           0 \\
           0 \\
           \bar{\alpha} \sqrt{\Delta}
     \end{pmatrix}, \label{eq:g_l}
 \end{equation}
$\vec{w}=(\bar{\alpha},\bar{\beta})^\text{t}$ is either $(1,0)^\text{t}$ or $(0,1)^\text{t}$ for $s = 1/2$ and $s=-1/2$, respectively, $\Delta=(1-c^2/\varepsilon)\sin^2 \theta_0$, $\varepsilon = c \sqrt{c^2+p^2_{\perp}+p^2_{\parallel}}$, $p_{\parallel}=|\boldsymbol{p}|\cos{\theta_0}$, and $p_{\perp}=|\boldsymbol{p}|\sin \theta_0$. The functions $a_k$ depend on $p_\parallel$, $p_\perp$, and $s$ but are independent of the spatial coordinates. The transverse coordinates can either be $x$ and $y$ or $\rho$ and $\varphi$. 
The wavefunction is determined by the following quantum numbers: $l$, $s$, $p_{\parallel}$, and $p_{\perp}$. 
Note that the orbital angular momentum and the spin projection are not well defined and only the projection of the total angular momentum $m = l+s$ is conserved. Note also that the Bessel beams with given $s$ have a nonzero projection onto the Volkov solutions~\eqref{eq:volkov_1} with the other sign of $s'$. The Bessel beams~\eqref{eq:bessel_init} are normalized according to
\begin{equation}
    \langle \psi^{(\text{B})}_{l,p_{\parallel},p_{\perp},s} | \psi^{(\text{B})}_{l,p_{\parallel}',p_{\perp}',s'} \rangle = \delta_{s s'} \delta (p_\parallel - p_\parallel') \delta (p_\perp - p_\perp'). \label{eq:Bessel_norm}
\end{equation}
This inner product is evaluated in Appendix~A.

In what follows, we will calculate the coefficients $C_{\boldsymbol{p}',s'}^{(\zeta)}$ [see Eq.~\eqref{eq:coeff_C}] using the explicit form of Volkov states [see Eqs.~(\ref{eq:volkov_1})--(\ref{eq:volkov_f_specified})]. One obtains
\begin{equation}
    C_{\boldsymbol{p}',s'}^{(\zeta)} = \frac{\sqrt{p_\perp}}{8\pi^{5/2}} \int\limits_{-\infty}^{\infty} \! dz \, \mathrm{e}^{i(p_{\parallel} - \zeta p'_{\parallel})z}\big [f_{\boldsymbol{p'},s'}^{(\zeta)}(t_{\rm in},z) \big ]^\dagger \int\limits_{-\infty}^{\infty} \! dx \! \int\limits_{-\infty}^{\infty} \! dy \, \mathrm{e}^{-i\zeta(p'_x x + p'_y y )} g_{l,p_{\parallel},p_{\perp},s}(x,y). \label{eq:C_twiste_gen}
\end{equation}
The double integral over $x$ and $y$ can be calculated analytically (see Appendix~A). We arrive at
\begin{equation}
C_{\boldsymbol{p}',s'}^{(\zeta)} = \delta(p_{\perp}-p'_{\perp}) c_{p_\parallel', \varphi_{p'},s'}^{(\zeta)},
\label{eq:C_delta_c}
\end{equation}
where $\boldsymbol{p}' = (p_\perp' \cos \varphi_{p'}, p_\perp' \sin \varphi_{p'}, p_\parallel')$ and
\begin{equation}
    c_{p_\parallel', \varphi_{p'},s'}^{(\zeta)} =\frac{1}{4\pi^{3/2}} \frac{1}{\sqrt{p_\perp}} \, i^{-\zeta l} \mathrm{e}^{il\varphi_{p'}} \int\limits_{-\infty}^{\infty} \! dz \, \mathrm{e}^{i(p_{\parallel} - \zeta p'_{\parallel})z} \big [f_{\boldsymbol{p}',s'}^{(\zeta)}(t_{\rm in},z) \big ]^\dagger \sum_{k=-1}^{+1} (-1)^{k \delta_{\zeta,-1}} a_k \mathrm{e}^{i k \varphi_{p'}}. \label{eq:c_bessel_final}
\end{equation}
Here $\delta_{\zeta,-1}$ reflects that the Bessel beam contains only the positive-energy components. Accordingly, we construct a decomposition of the Bessel beam in terms of plane waves with given $p_\perp$ which are determined by the momentum projection $p_\parallel'$, direction of the transverse momentum component $\varphi_{p'}$, and spin quantum number $s$. 
In what follows, we will discuss the analogous expansion in the case of a localized twisted WP.

\subsection{Twisted wavepacket}\label{sec:twisted_packet}
To make the wavefunction finite in the $z$ direction, one has to combine solutions with different longitudinal momenta. In the case of a plane-wave electron, these solutions usually correspond to given values of the transverse projections $p_x$ and $p_y$. However, in the case of vortex states, these are not well defined. In this study, we will fix the magnitude of the transverse momentum component $p_\perp$, so the expression for the twisted WP reads
\begin{equation}\label{eq:bessel_init_tw}
    \psi^{(0)}_{l,p_{\parallel},p_\perp,s}(\boldsymbol{x}) = \int \limits_{-\infty}^{+\infty} \! dq \, f(q) 
    \psi^{(\text{B})}_{l,p_{\parallel} + q,p_{\perp},s}(\boldsymbol{x}), 
\end{equation}
where the smearing function $f(q)$ satisfies 
\begin{equation}\label{eq:f_norm}
    \int \limits_{-\infty}^{+\infty} \! d q \, |f(q)|^2 = 1.
\end{equation}
We choose a Gaussian profile
\begin{equation}
    f(q) = \frac{1}{(\pi \sigma^2)^{1/4}}
     \exp \left(- q^2 / 2 \sigma^2 \right),
\label{eq:f_Gaussian}
\end{equation}
where $\sigma$ determines the WP width in momentum space, i.e. the uncertainty in $p_\parallel$. In all our calculations we set $\sigma = $~10~a.u. 
%

\subsection{Plane-wave decomposition of a twisted wavepacket} \label{twisted_packet_decomposition}

Assuming that at $t = t_\text{in} = -(L+ \xi_\text{max})/c$ the WP has the form~\eqref{eq:bessel_init_tw}, we will now evaluate the positive-energy coefficients with the aid of Eqs.~\eqref{eq:C_delta_c} and \eqref{eq:c_bessel_final}. Since the laser field almost vanishes for $\xi < -\xi_\text{max}$, its left edge is initially located to the right of $z = L$. We assume that the electron WP is localized within the interval $[-L,\ L]$. It means that in Eq.~(\ref{eq:c_bessel_final}), where $p_\parallel$ should be replaced with $p_\parallel + q$ according to Eq.~\eqref{eq:bessel_init_tw}, one can integrate only over this region. The condition $z \in [-L,\ L]$ corresponds to $\xi \in [-2L -\xi_\text{max} ,\ -\xi_\text{max}]$, so the external field and its potential involved in $f_{\boldsymbol{p}',s'}^{(\zeta)}(t_{\rm in},z)$ vanish. It allows us to perform integration over $z$:
\begin{equation}
\int\limits_{-L}^{L} \! dz \, \mathrm{e}^{i(p_{\parallel} + q - p'_{\parallel})z} = \frac{2 \sin(p_\parallel + q - p_\parallel')L}{p_\parallel + q - p_\parallel'}.
\label{eq:z_integration}
\end{equation}
The results proved to be independent of $L$ and $\xi_\text{max}$, provided they are sufficiently large. One obtains
\begin{equation}
C_{\boldsymbol{p}',s'}^{(+)} = \frac{1}{2 \pi^{3/2}} \, i^{-l} \mathrm{e}^{i l \varphi_{p'}} \int \limits_{-\infty}^{+\infty} dq \, f(q) \, \mathrm{e}^{i \varepsilon' t_\text{in}} (p_\perp)^{-1/2} \delta (p_\perp - p_\perp') \, \frac{\sin(p_\parallel + q - p_\parallel')L}{p_\parallel + q - p_\parallel'} \, \sum_{k=-1}^{+1} \mathrm{e}^{i k \varphi_{p'}} u^\dagger (\boldsymbol{p}', s') a_k.
\label{eq:C_packet_final}
\end{equation}
Here $A_k$ involve $p_\parallel + q$. The wavefunction at arbitrary time instant $t$ can then be evaluated via
\begin{equation}
\psi_{\boldsymbol{p}, s} (t, \boldsymbol{x}) = \frac{i^{-l}\sqrt{p_\perp}}{2 \pi^{3/2}} \sum_{s'} \int \limits_{-\infty}^{+\infty} \! dp_\parallel' \int \limits_0^{2\pi} \! d \varphi_{p'} \, \mathrm{e}^{i l \varphi_{p'}} \int \limits_{-\infty}^{+\infty} dq f(q) \, \mathrm{e}^{i \varepsilon' t_\text{in}} \, \frac{\sin(p_\parallel + q - p_\parallel')L}{p_\parallel + q - p_\parallel'} \sum_{k=-1}^{+1} \mathrm{e}^{i k \varphi_{p'}} [ u^\dagger (\boldsymbol{p}', s') a_k ] \varphi^{(+)}_{\boldsymbol{p}', s'} (t, \boldsymbol{x}).
\label{eq:psi_t_tw}
\end{equation}
In this expression, $\boldsymbol{p}' = (p_\perp \cos \varphi_{p'}, p_\perp \sin \varphi_{p'}, p_\parallel')$.

\end{widetext}

\section{Results}\label{sec:results}

Here we implement the procedure described in Sec.~\ref{sec:twisted} and examine various observable quantities concerning the structure and dynamics of twisted WPs interacting with finite laser pulses. We start our analysis from the following set of parameters. The laser field parameters are $I \approx 1.3 \times 10^{13}$ W/cm$^2$, $a = 9$, and $\omega = 0.242$ a.u. The electron state corresponds to $l = 3$, $\theta_0 = \pi / 4$, and kinetic energy $E_{\rm kin} = 817.4$~keV. These parameters are chosen as in Ref.~\cite{HayrapetyanPRL2014}. The WP width in momentum space is $\sigma = $ 10~a.u. Note that in Ref.~\cite{HayrapetyanPRL2014} the Gaussian envelope function is involved in the vector potential, while here we use it in the electric field itself [see Eq.~\eqref{eq:field}], which allows us to construct unipolar laser pulses, $\mathcal{A}_0 \neq 0$.

\subsection{Overall dynamics}

\begin{figure*}[t]
\center{\includegraphics[width=0.9\linewidth]{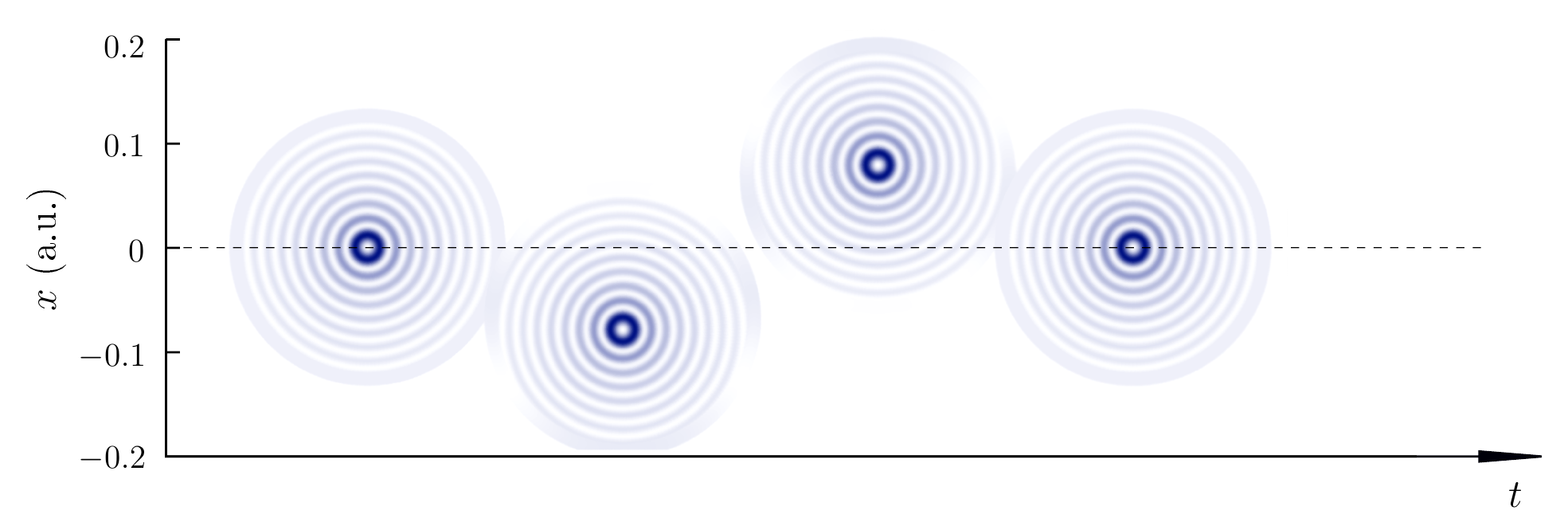}}
    \caption{Snapshots of the twisted WP density integrated over $z$ at different $t$. The $x$ coordinate corresponds to the position of the WP center. The laser field and WP parameters are $I \approx 1.3 \times 10^{13}$ W/cm$^2$, $a = 9$, $\omega = 0.242$ a.u., $l = 3$, $\theta_0 = \pi / 4$, and $E_{\rm kin} = 817.4$~keV.}
    \label{fig:movie_ordinary}
\end{figure*}

\begin{figure*}[t]
\center{\includegraphics[width=0.9\linewidth]{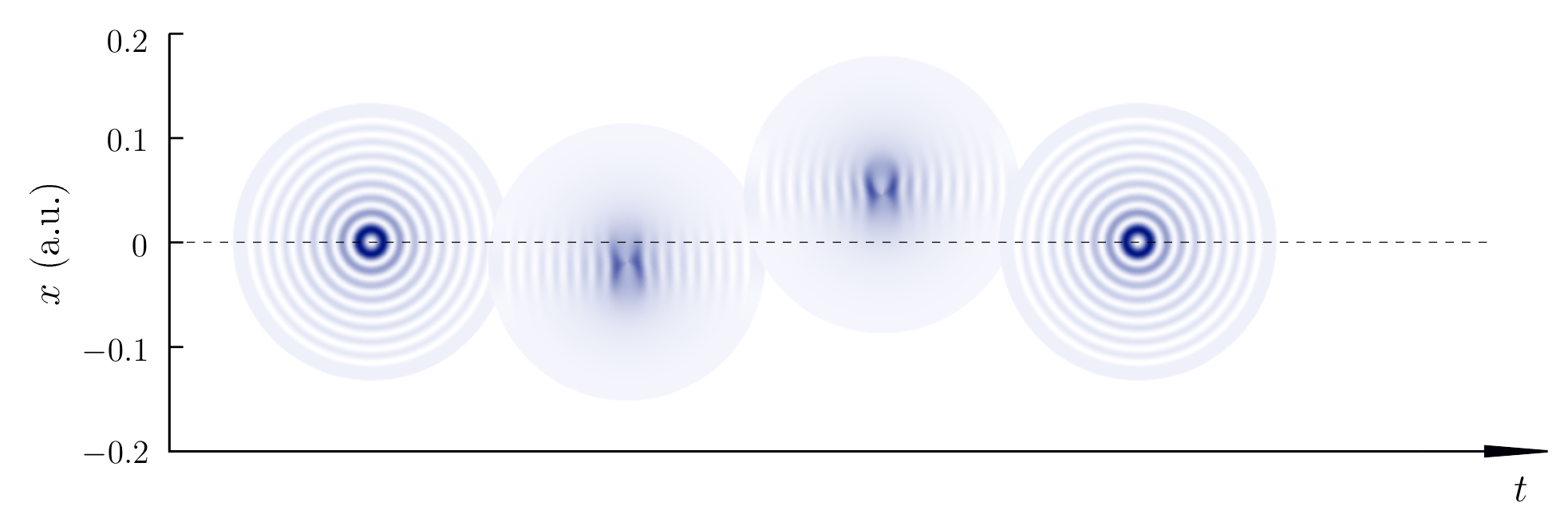}}
    \caption{Snapshots of the twisted WP density integrated over $z$ at different $t$. The $x$ coordinate corresponds to the position of the WP center. The laser frequency is $\omega = 4.84$ a.u. and intensity is $I \approx 2.1 \times 10^{18}$ W/cm$^2$. The rest parameters are the same as in Fig.~\ref{fig:movie_ordinary}.}
    \label{fig:movie_zusammenbruch}
\end{figure*}

In order to analyze the twisted WP dynamics, we integrate the electron probability density $|\psi_{\boldsymbol{p}, s} (t, \boldsymbol{x})|^2$ over $z$ at various time instants $t$ and plot the corresponding snapshots in the $xy$ plane. As the external field does not exert a force along the $y$ axis, we observe displacements only in the $x$ direction. Several snapshots are displayed in Fig.~\ref{fig:movie_ordinary}. Although the WP moves as a whole along the $x$ axis, its ring structure remains the same~\cite{HayrapetyanPRL2014}. However, this is not always the case as demonstrated in the following example. In Fig.~\ref{fig:movie_zusammenbruch} we present the analogous results for higher frequency and amplitude of the laser pulse: $\omega = 4.84$ a.u. and $I \approx 2.1 \times 10^{18}$ W/cm$^2$. In this case, the spatial extent of the twisted WP exceeds the laser wavelength, so different ``slices'' perpendicular to the $z$ axis interact with different parts of the laser pulse and thus have different $x$ displacements. Accordingly, the overall structure of the WP blurs or even disintegrates. As will be discussed later, this effect leads also to large uncertainties in the angular momentum of the twisted electron. 
Note that the parameters chosen in Figs.~\ref{fig:movie_ordinary} and \ref{fig:movie_zusammenbruch} lead to the same value of the classical nonrelativistic oscillation amplitude, so the $x$ displacements do not differ much for these two field configurations.

We have not yet explored the WP dynamics with respect to the $z$ axis, nor have we specified the $x(t)$ dependence. It turns out that great insights can be obtained from the analysis of classical trajectories. This will be discussed next.

\subsection{Comparison with classical motion}

Twisted electron states do not possess well-defined momentum projections along the $x$ and $y$ axes as they are characterized only by the magnitude $p_\perp$ of the transverse momentum component. It suggests that the dynamics of a twisted electron cannot be traced by solving the classical equations of motion. Here we will evolve classical trajectories with various initial directions of the transverse momentum component and compare them with the results of our quantum simulations based on solving the Dirac equation. As will be seen, the crucial parameter here is the electric-field area of the laser pulse:
\begin{equation}
S_E = \int \limits_{-\infty}^{+\infty} E_x (t,z) dt = -\frac{1}{c} \, \mathcal{A}_0.
\label{eq:field_area_SE}
\end{equation}
The classical equations of motion can be solved by following the same procedure as was outlined in Ref.~\cite{Aleksandrov_2020}, where the external field was assumed to be monochromatic. Let $\boldsymbol{p}_0$ be the initial momentum of the electron. Then the final momentum projections read
\begin{eqnarray}
p_x &=& p_{0x} - S_E, \label{eq:class_px}\\
p_y &=& p_{0y}, \label{eq:class_py}\\
p_z &=& p_{0z} - \frac{S_E}{\varepsilon_0/c - p_{0z}} \, \Big ( p_{0x} + \frac{1}{2}\, S_E \Big ), \label{eq:class_pz}
\end{eqnarray}
where $\varepsilon_0 = c \sqrt{c^2 + \boldsymbol{p}_0^2}$. Although the $z$ component of the momentum of the twisted WP~\eqref{eq:bessel_init_tw} approximately corresponds to $p_\parallel$, the $x$ and $y$ projections are completely unknown and only obey $p_{0x}^2 + p_{0y}^2 = p_\perp^2$. Let us introduce angle $\varphi_{p_0}$: $\boldsymbol{p}_0 = (p_\perp \cos \varphi_{p_0}, p_\perp \sin \varphi_{p_0}, p_\parallel)$. Changing $\varphi_{p_0}$ we obtain a set of classical trajectories, none of which is supposed to follow the mean values of the momentum and coordinate operators within the quantum computations. However, we will also average the results over $\varphi_{p_0}$ in order to approximately predict the quantum behavior. Our goal is to examine the accuracy of this approach although it completely neglects the effects of quantum interference.

\begin{figure}[t]
\center{\includegraphics[width=0.98\linewidth]{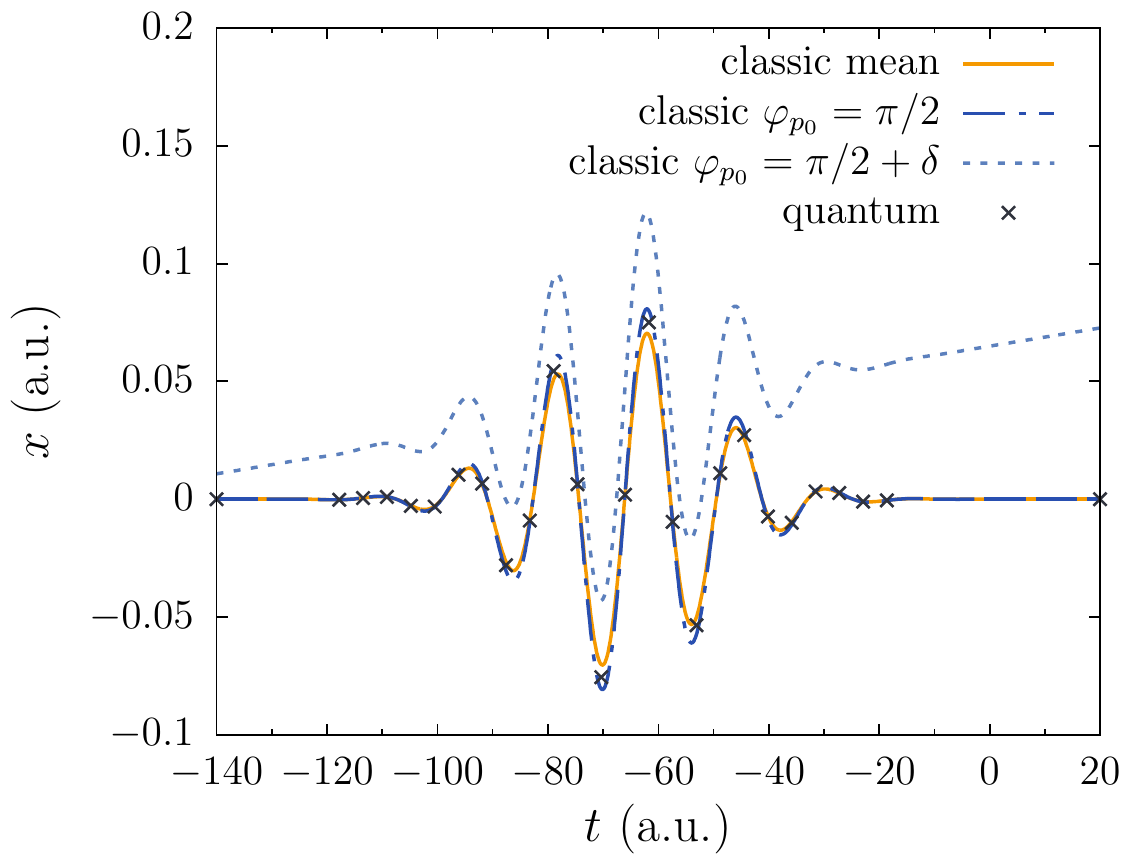}}
    \caption{Comparison between the mean value of the $x$ coordinate of the twisted WP and the classical predictions for various directions $\varphi_{p_0}$ of the initial transverse momentum. Here $\delta = -4 \times 10^{-6}$. The electron and laser parameters are the same as in Fig.~\ref{fig:movie_ordinary}.}
    \label{fig:xt}
\end{figure}

In Fig.~\ref{fig:xt} we display the mean values of the $x$ coordinate of the electron WP as a function of $t$ for the parameters used in Fig.~\ref{fig:movie_ordinary}. One observes that by averaging the classical predictions over $\varphi_{p_0}$, one obtains a very accurate approximation of the results of quantum simulations. Although the classical trajectories are extremely sensitive to the value of $\varphi_{p_0}$, for $\varphi_{p_0} = \pi/2$ the results are also close to the quantum mean. Note, however, that in the case $\varphi_{p_0} = \pi/2$, the $y$ coordinate tends to infinity with increasing $t$, while the actual position of the electron WP always corresponds to $y=0$. This suggests that one has indeed to average the classical predictions. Moreover, the classical trajectories turn out to be extremely sensitive to the value of $\varphi_{p_0}$ (see the results for $\varphi_{p_0} = \pi/2 + \delta$ in Fig.~\ref{fig:xt}).

\begin{figure}[t]
\center{\includegraphics[width=0.98\linewidth]{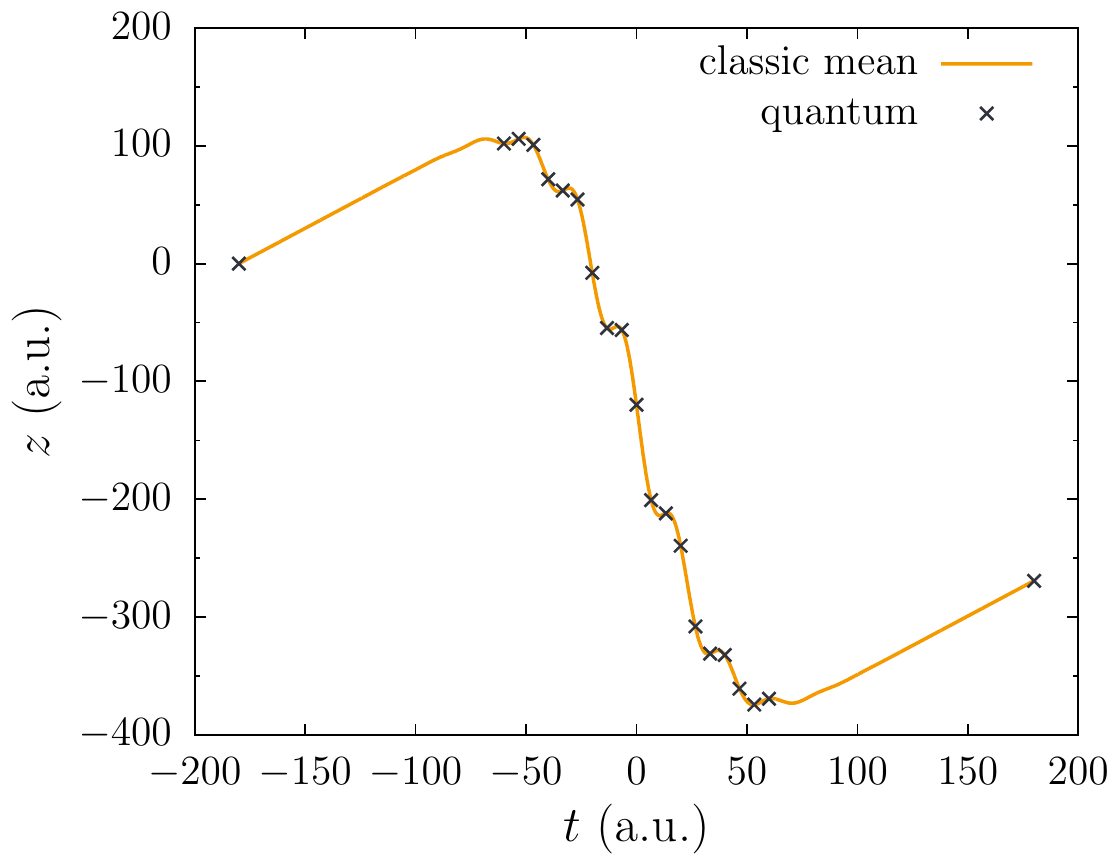}}
    \caption{Comparison between the mean value of the $z$ coordinate of the twisted WP and the classical predictions averaged over the direction $\varphi_{p_0}$ of the initial transverse momentum. The laser field and WP parameters are $I \approx 3.5 \times 10^{18}$ W/cm$^2$, $a = 9$, $\omega = 0.15$ a.u., $l = 3$, $\theta_0 = 11.3^{\circ}$,  and $E_{\rm kin} = 0.014$~keV.}
    \label{fig:zt}
\end{figure}

In Fig.~\ref{fig:zt} we show that the mean classical trajectory is also very close to the quantum mean of the $z$ coordinate. Here we use different parameters to make the curve nontrivial and thus more evidently demonstrate the efficiency of the classical approach. Finally, we note that classical trajectories are also used to guess a proper position of the spatial box within quantum calculations.

\begin{figure}[t]
\includegraphics[width=0.98\linewidth]{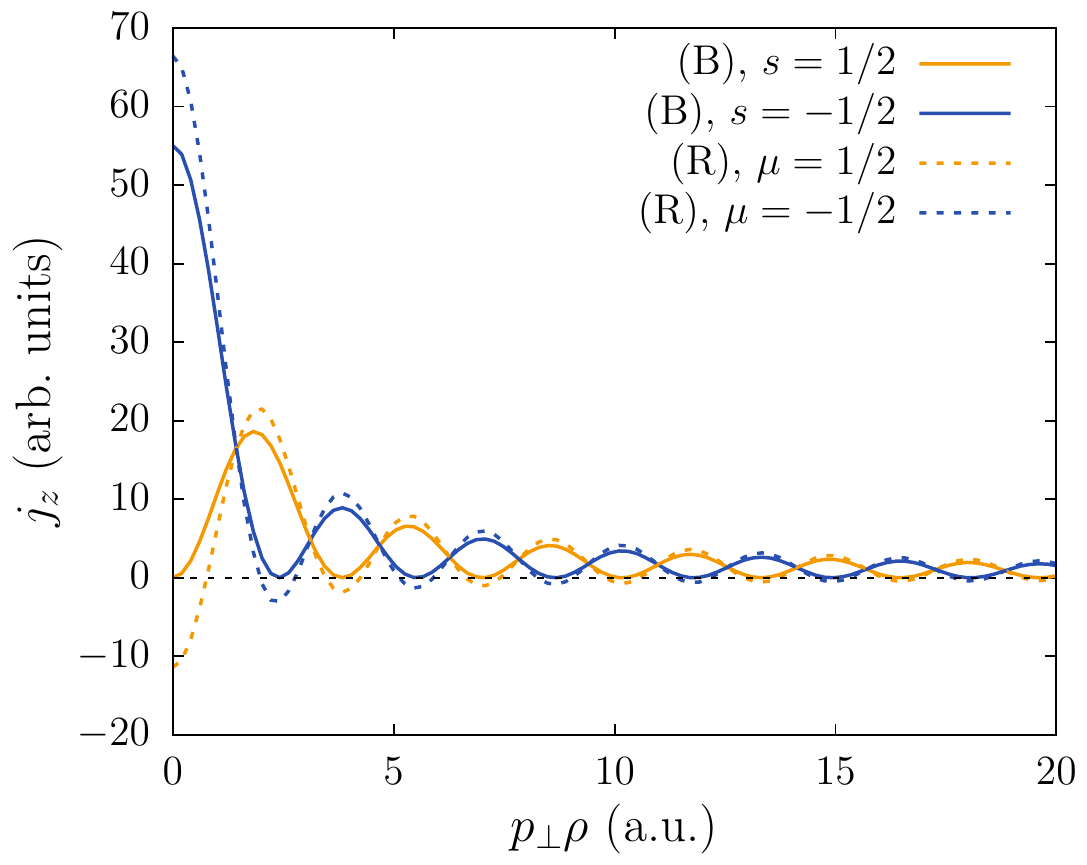}
    \caption{Local current density of a twisted electron integrated over the polar angle $\varphi$ and $z$ coordinate at $t=t_\text{in}$. The electron state with $m = l + s = - 1/2$ is constructed from the Bessel beams~\eqref{eq:bessel_init} [labeled as (B), solid lines] and from the ``rotated'' functions~\eqref{eq:bessel_rot_1} and \eqref{eq:bessel_rot_2} [labeled as (R), dashed lines]. In the latter case, the current density can have negative values. The momentum components are $p_\perp = p_\parallel$ = 10 a.u.}
    \label{fig:current}
\end{figure}

\subsection{Current density}

 Now we turn to the analysis of the local current density along the $z$ axis in the absence of the external field. As was demonstrated in Refs.~\cite{Kosheleva_2018,Groshev_2020}, in the case of Bessel beams, this quantity may be negative at certain positions in space, while the wavepacket itself travels with a positive speed in the $z$ direction (previously, the analogous finding in the case of twisted light was reported in Ref.~\cite{surzhykov_pra_2016}). However, the local current values in the states~\eqref{eq:bessel_init} are given by
\begin{eqnarray}
j_z (\boldsymbol{x}) &=& \psi^{(\text{B})\dagger}_{l,p_{\parallel},p_{\perp},s} (\boldsymbol{x}) \alpha_z \psi^{(\text{B})}_{l,p_{\parallel},p_{\perp},s} (\boldsymbol{x}) \nonumber \\
{} &=& \frac{p_\perp}{4\pi^2} \frac{p c}{\varepsilon} \cos \theta_0 J^2_l (p_\perp \rho),
\label{eq:current_bessel}
\end{eqnarray}
where $p = \sqrt{p_\perp^2 + p_\parallel^2}$, so the current is always positive. It turns out that one may indeed obtain negative values of $j_z$ if one constructs a linear combination of two Bessel beams~\eqref{eq:bessel_init} with given $m = l+s$ and $s = \pm 1/2$. The functions used in Refs.~\cite{Kosheleva_2018,Groshev_2020} are exactly those combinations (see Appendix~B).

Let us construct wavefunctions $\psi^{(\text{R})}_{m,p_{\parallel},p_{\perp},\mu}$ according to
\begin{eqnarray}
\psi^{(\text{R})}_{m,p_{\parallel},p_{\perp},1/2} (\boldsymbol{x}) &=& \cos \frac{\theta_0}{2} \psi^{(\text{B})}_{m-1/2,p_{\parallel},p_{\perp},1/2} (\boldsymbol{x}) \nonumber \\
{} &+& i \sin \frac{\theta_0}{2} \psi^{(\text{B})}_{m+1/2,p_{\parallel},p_{\perp},-1/2} (\boldsymbol{x}),
\label{eq:bessel_rot_1} \\
\psi^{(\text{R})}_{m,p_{\parallel},p_{\perp},-1/2} (\boldsymbol{x}) &=& i \sin \frac{\theta_0}{2} \psi^{(\text{B})}_{m-1/2,p_{\parallel},p_{\perp},1/2} (\boldsymbol{x}) \nonumber \\
{} &+& \cos \frac{\theta_0}{2} \psi^{(\text{B})}_{m+1/2,p_{\parallel},p_{\perp},-1/2} (\boldsymbol{x}).
\label{eq:bessel_rot_2}
\end{eqnarray}
We distinguish between the functions~\eqref{eq:bessel_rot_1} and \eqref{eq:bessel_rot_2} by means of the quantum number $\mu$, which determines the helicity of the electron beam~\cite{serbo_pra_2015}. Instead of Eq.~\eqref{eq:current_bessel}, it yields now
\begin{eqnarray}
\tilde{j}_z (\boldsymbol{x}) &=& \frac{p_\perp}{4\pi^2} \frac{p c}{\varepsilon} \bigg \{ \big [\cos (\theta_0/2) J_{m-\mu} (p_\perp \rho) \big ]^2 \nonumber\\
{} &-& \big [\sin (\theta_0/2) J_{m+\mu} (p_\perp \rho)\big ]^2 \bigg \},
\label{eq:current_bessel_new}
\end{eqnarray}
which coincides, e.g., with Eq.~(13) from Ref.~\cite{Kosheleva_2018} up to a common normalization factor $1/(2\pi p_\perp)$. We perform computations with the twisted WP~\eqref{eq:bessel_init_tw} consisting of the Bessel beams~\eqref{eq:bessel_init} and with a similar combination of the ``rotated'' beams~\eqref{eq:bessel_rot_1} and \eqref{eq:bessel_rot_2}. The results for $p_\perp = p_\parallel$ = 10 a.u., and $m = -1/2$ at $t = t_\text{in}$ are given in Fig.~\ref{fig:current}, where we present the local current density integrated over the polar angle $\varphi$ and over the $z$ coordinate. We observe that the local current density can indeed be negative in the case of the ``rotated'' wavefunctions (dashed lines). Finally, we note that the sum of two currents in Eq.~\eqref{eq:current_bessel_new} with different signs of $\mu$ precisely corresponds to the sum of currents~\eqref{eq:current_bessel} for given $m$ and $l = m \pm 1/2$, i.e., the sum of the solid lines in Fig.~\ref{fig:current} equals the sum of the dashed ones.

\begin{figure*}[t]
\center{\includegraphics[height=0.37\linewidth]{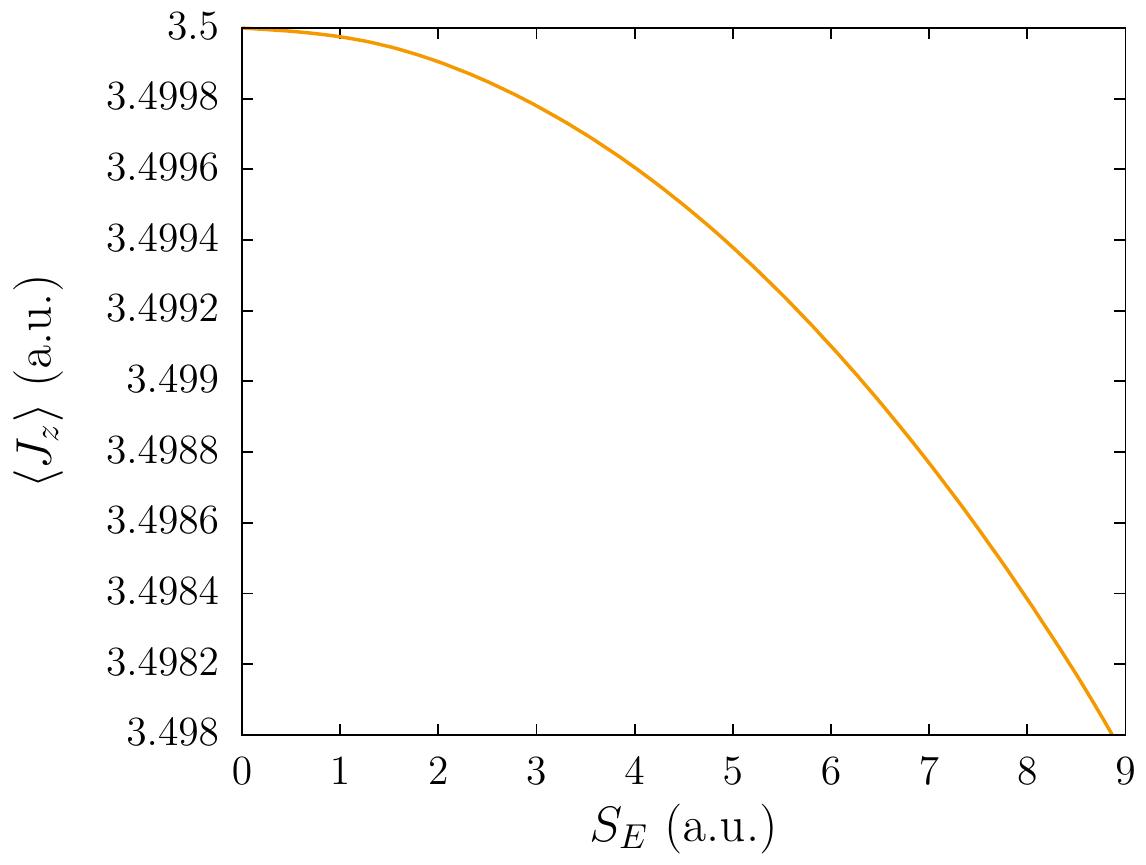}~~~~~\includegraphics[height=0.37\linewidth]{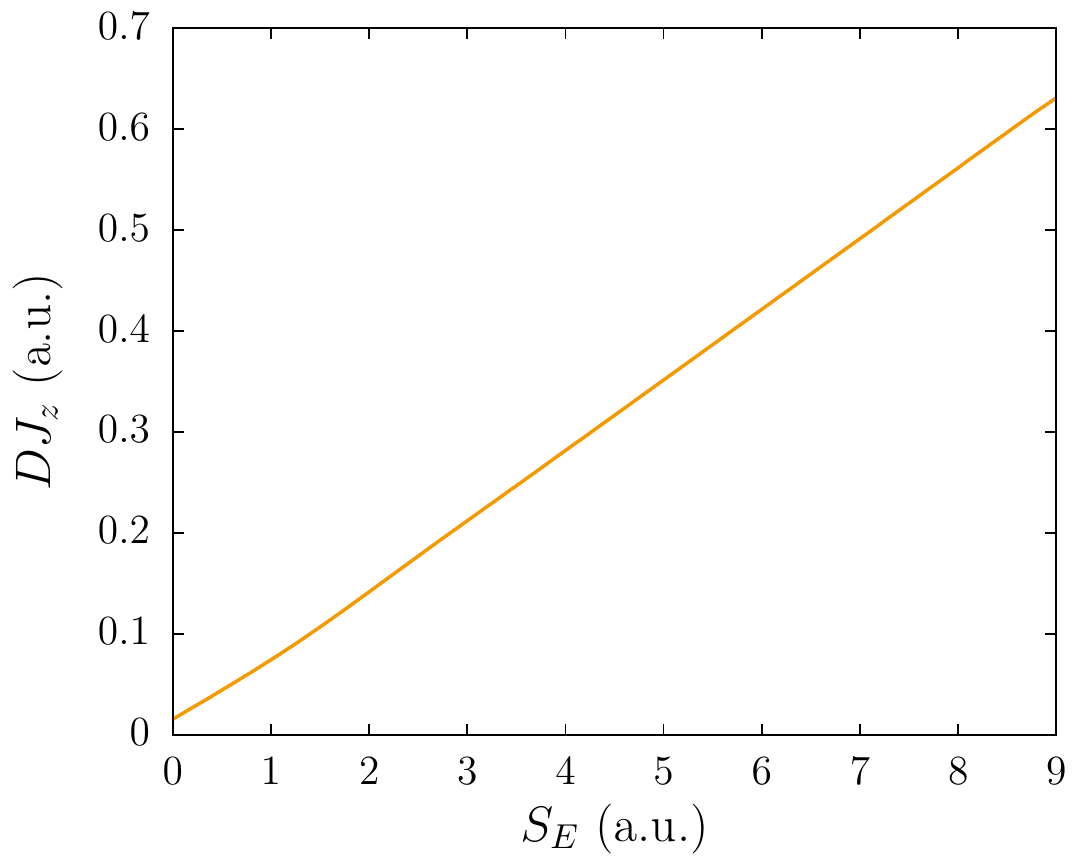}}
    \caption{Mean values of the $z$ projection of the total angular momentum as a function of the field area $S_E$ (left) and the dispersion $DJ_z$ (right). The parameters are: $I \approx 3.5 \times 10^{16}$ W/cm$^2$, $a=0.9$, $\omega = 0.15$ a.u., $l = 3$, $\theta_0 = 11.3^{\circ}$, and $E_{\rm kin} = 1.41$ keV.}
    \label{fig:am}
\end{figure*}

\subsection{Total angular momentum and its dispersion}

The twisted WP~\eqref{eq:bessel_init_tw} involves Bessel beams with given numbers $l$ and $s$, which means that the electron wavefunction has initially a well-defined value of the $z$ projection of the total angular momentum $m=l+s$. Here we will evaluate the uncertainty of this number which arises due to the interaction with the external laser field. To this end, we will compute a mean value of the operator $J_z = L_z + S_z$ and its dispersion (standard deviation),
\begin{equation}
DJ_z = \sqrt{\langle J^2_z \rangle - \langle J_z \rangle^2}.
\end{equation}
In the case of an individual Bessel beam $\psi^{(\text{B})}_{l,p_{\parallel},p_{\perp},s}$,  one obtains
\begin{eqnarray}
\langle L_z \rangle &=& l + s(1-c^2/\varepsilon) \sin^2 \theta_0, \\
\langle S_z \rangle &=& s - s(1-c^2/\varepsilon) \sin^2 \theta_0, \\
\langle J_z \rangle &=& l + s = m, \\
DJ_z &=& 0.
\end{eqnarray}
Combining Bessel beams according to Eq.~\eqref{eq:bessel_init_tw} and evolving the WP in time, we find out that the interaction with the laser pulse can lead to large values of $DJ_z$. In Fig.~\ref{fig:am} we present $\langle J_z \rangle$ and $DJ_z$ at the final time instant $t = t_\text{out}$ as a function of the electric-field area $S_E$ for the following parameters: 
$I \approx 3.5 \times 10^{16}$ W/cm$^2$, $a=0.9$, $\omega = 0.15$ a.u., $l = 3$, $\theta_0 = 11.3^{\circ}$, and $E_{\rm kin} = 1.41$~keV. We change the field area by means of the CEP parameter $\phi$ [see Eq.~\eqref{eq:field_area}]. The mean value of the total angular momentum remains almost the same even for large $S_E$ but there is still a certain deviation from $3.5$~a.u. due to the changes in the spin projection $\langle S_z \rangle$~\cite{Aleksandrov_2020}. More important, the dispersion turns out to be significant. This effect can be explained if one recalls that the $z$ slices of the electron wavefunction can have very different displacements along the $x$ direction. While it does not alter the mean value $\langle J_z \rangle$ since $\langle y \rangle = 0$ and $\langle p_y \rangle = 0$, the dispersion $DJ_z$ increases with the relative displacement among different $z$ slices, which, in turn, is proportional to $S_E$. This suggests that a highly unipolar finite laser pulse can substantially damage the vortex structure of the electron state.

\section{Conclusion}\label{sec:conclusion}

In this study, the interaction of a twisted wavepacket with a linearly polarized laser pulse was investigated within relativistic quantum mechanics. The process was simulated taking into account the finite size of the both objects. A special focus was placed on the dynamics of the electron state and evolution of its various characteristics. It was demonstrated that the coordinates of the center of the wavepacket can be accurately determined by averaging over the classical trajectories corresponding to different orientations of the initial momentum of the particle. This observation is expected to be highly beneficial as it allows one to obtain valuable predictions without performing quantum calculations, i.e., without solving the Dirac equation. The analysis of classical trajectories was also extremely helpful in interpreting other effects revealed in our study. Namely, it was shown that the vortex structure of the electronic state can be destroyed by highly unipolar laser pulses. This was found by a direct inspection of the ring structure of the wavepacket and also by computing the dispersion of the total angular momentum, which proved to rapidly increase with the pulse area. Besides, it was demonstrated that the local current density can have negative values if the Bessel beams are properly combined within the electron wavefunction.



\appendix

\section{Normalization of Bessel beams and integration over transverse coordinates}

Here we will show that the Bessel beams~(\ref{eq:bessel_init}) are normalized according to Eq.~\eqref{eq:Bessel_norm}. Although the presence of $\delta(p_{\parallel}-p'_{\parallel})$ in Eq.~(\ref{eq:Bessel_norm}) is obvious, let us briefly discuss how the second delta-function should appear. According to Eq.~(\ref{eq:g_l}), the wavefunction contains three terms with Bessel functions of different orders. First, we note that the integration over $\varphi$ makes all of the terms with two different orders vanish, so one has to consider only ``diagonal terms''. The corresponding integral over $\rho$ reads
\begin{equation}
    \int\limits_0^{\infty} \! d\rho \rho J_{l}(p_{\perp}\rho)J_l(p'_{\perp}\rho).
\end{equation}
Calculating this integral with a finite upper limit, one can show that
\begin{equation}
    \int\limits_0^{R} \! d\rho \rho J_{l}(p_{\perp}\rho)J_l(p'_{\perp}\rho)\underset{R\rightarrow\infty}{\longrightarrow}\frac{1}{p_\perp}\delta(p_{\perp}-p'_{\perp}),
\label{eq:JJ}
\end{equation}
provided $p_{\perp},p'_{\perp}>0$. Combining all the terms, we obtain Eq.~(\ref{eq:Bessel_norm}).

The relation~(\ref{eq:JJ}) is also useful in calculating the integral over $x$ and $y$ in Eq.~(\ref{eq:C_twiste_gen}). According to Eq.~(\ref{eq:g_AAA}), there are three terms. Let us consider that containing $a_k$:
\begin{eqnarray}
\mathcal{I}_k & \equiv & a_k i^k \int\limits_{-\infty}^{\infty} \! dx \! \int\limits_{-\infty}^{\infty} \! dy \,  \mathrm{e}^{-i\zeta(p'_x x + p'_y y )}\mathrm{e}^{i(l+k) \varphi}J_{l+k}(p_{\perp}\rho) \nonumber \\ 
    {}&=&a_k i^k \int\limits_0^\infty \! d\rho \rho \int\limits_0^{2\pi} \! d\varphi \, \mathrm{e}^{[-i\zeta p'_{\perp}\rho\cos{(\varphi_{p'}-\varphi)}]} \nonumber \\
    {}&\times& \mathrm{e}^{i(l+k) \varphi}J_{l+k}(p_{\perp}\rho).
\end{eqnarray}
To perform integration over $\varphi$, we utilize the following identity~\cite{stegun}:
\begin{equation}
\int\limits_0^{2\pi} \! d\varphi \,  \mathrm{e}^{iz \cos \varphi} \, \mathrm{e}^{in\varphi} = 2 \pi i^n J_n (z).
\end{equation}
Thus, one obtains
\begin{eqnarray}
\mathcal{I}_k & = & 2\pi a_k i^{2k+l} \mathrm{e}^{i(l+k)\varphi_{p'}} \int\limits_0^{\infty} \! d\rho \rho J_{l+k} (p_{\perp}\rho)J_{l+k}(-\zeta p'_{\perp}\rho) \nonumber \\
    {} & = & \frac{2\pi a_k}{p_\perp} \, (-1)^{k\delta_{\zeta,-1}} i^{-\zeta l} \mathrm{e}^{i(l+k)\varphi_{p'}} \,  \delta(p_{\perp}-p'_{\perp}).
\label{eq:I_app}
\end{eqnarray}
Here, we have used $J_l (-z) = (-1)^l J_l (z)$. It immediately brings us to Eq.~(\ref{eq:c_bessel_final}).

\begin{widetext}
\section{Connection with the wavefunctions from Refs.~\cite{serbo_pra_2015,Kosheleva_2018,Groshev_2020}}

In Refs.~\cite{Kosheleva_2018,Groshev_2020} the relativistic wavefunctions of twisted electrons were chosen in the form $\tilde{\psi}_{m,p_{\parallel},p_{\perp},\mu}$,
\begin{eqnarray}
\tilde{\psi}_{m,p_{\parallel},p_{\perp},1/2} (\boldsymbol{x}) &=& \frac{1}{(2\pi)^{3/2}} \frac{\mathrm{e}^{ip_{\parallel}z}}{\sqrt{2\varepsilon}}
\begin{pmatrix}
           \sqrt{\varepsilon + c^2} \cos (\theta_0/2) \, \mathrm{e}^{i(m-1/2)\varphi} J_{m-1/2} (p_\perp \rho)\\
           i\sqrt{\varepsilon + c^2} \sin (\theta_0/2) \, \mathrm{e}^{i(m+1/2)\varphi} J_{m+1/2} (p_\perp \rho) \\
           \sqrt{\varepsilon - c^2} \cos (\theta_0/2) \, \mathrm{e}^{i(m-1/2)\varphi} J_{m-1/2} (p_\perp \rho) \\
           i\sqrt{\varepsilon - c^2} \sin (\theta_0/2) \, \mathrm{e}^{i(m+1/2)\varphi} J_{m+1/2} (p_\perp \rho)
     \end{pmatrix}, \label{eq:zaytsev_wf_1} \\
\tilde{\psi}_{m,p_{\parallel},p_{\perp},-1/2} (\boldsymbol{x}) &=& \frac{1}{(2\pi)^{3/2}} \frac{\mathrm{e}^{ip_{\parallel}z}}{\sqrt{2\varepsilon}}
\begin{pmatrix}
           i\sqrt{\varepsilon + c^2} \sin (\theta_0/2) \, \mathrm{e}^{i(m-1/2)\varphi} J_{m-1/2} (p_\perp \rho) \\
           \sqrt{\varepsilon + c^2} \cos (\theta_0/2) \, \mathrm{e}^{i(m+1/2)\varphi} J_{m+1/2} (p_\perp \rho) \\
           -i\sqrt{\varepsilon - c^2} \sin (\theta_0/2) \, \mathrm{e}^{i(m-1/2)\varphi} J_{m-1/2} (p_\perp \rho) \\
           -\sqrt{\varepsilon - c^2} \cos (\theta_0/2) \, \mathrm{e}^{i(m+1/2)\varphi} J_{m+1/2} (p_\perp \rho)
     \end{pmatrix}, \label{eq:zaytsev_wf_2}
 \end{eqnarray}
where $m$ is the projection of the total angular momentum as used in the main text and $\mu$ is an additional quantum number governing the helicity of the beam~\cite{serbo_pra_2015}. These solutions are normalized according to
\begin{equation}
    \langle \tilde{\psi}_{m,p_{\parallel},p_{\perp},\mu} | \tilde{\psi}_{m,p_{\parallel}',p_{\perp}',\mu'} \rangle = \frac{1}{2\pi p_\perp} \, \delta_{\mu \mu'} \delta (p_\parallel - p_\parallel') \delta (p_\perp - p_\perp'). \label{eq:zaytsev_wf_norm}
\end{equation}
For a given value of $m$ and given momentum, Eq.~\eqref{eq:bessel_init} yields two functions with $s = \pm 1/2$ and $l = m-s$. Each of these functions can be represented as a combination of the two solutions~\eqref{eq:zaytsev_wf_1} and~\eqref{eq:zaytsev_wf_2}:
\begin{equation}
\begin{pmatrix}
          \psi^{(\text{B})}_{m-1/2,p_{\parallel},p_{\perp},1/2} \\
          \psi^{(\text{B})}_{m+1/2,p_{\parallel},p_{\perp},-1/2}
     \end{pmatrix} = \sqrt{2 \pi p_\perp}
     \begin{pmatrix}
         \cos (\theta_0/2) & -i\sin (\theta_0/2) \\
          -i\sin (\theta_0/2) & \cos (\theta_0/2)
     \end{pmatrix}
     \begin{pmatrix}
           \tilde{\psi}_{m,p_{\parallel},p_{\perp},1/2} \\
           \tilde{\psi}_{m,p_{\parallel},p_{\perp},-1/2}
     \end{pmatrix}.
     \label{eq:zaytsev_wf_connection}
\end{equation}
Accordingly, the wavefunctions~\eqref{eq:zaytsev_wf_1} and~\eqref{eq:zaytsev_wf_2} coincide with the functions~\eqref{eq:bessel_rot_1} and \eqref{eq:bessel_rot_2} up to a factor of $\sqrt{2 \pi p_\perp}$:
\begin{equation}
\tilde{\psi}_{m,p_{\parallel},p_{\perp},\mu} (\boldsymbol{x}) = \sqrt{2 \pi p_\perp} \psi^{(\text{R})}_{m,p_{\parallel},p_{\perp}, \mu } (\boldsymbol{x}). \label{eq:zaytsev_vs_rot}
\end{equation}

Finally, we note that in Ref.~\cite{serbo_pra_2015} twisted electron states were described by the wavefunctions which differ from those in Eqs.~\eqref{eq:zaytsev_wf_1} and~\eqref{eq:zaytsev_wf_2} only in an overall numerical factor.
\end{widetext}


%

\end{document}